\documentclass[reprint,prl,superscriptaddress]{revtex4-1}
\usepackage{graphicx}  
\usepackage{dcolumn}   
\usepackage{bm}        
\usepackage{amssymb}   
\usepackage{color}
\usepackage{xspace}
\usepackage{tikz}
\usepackage{pgffor}
\usepackage{color}
\usetikzlibrary{positioning}
\usetikzlibrary{calc}
\usetikzlibrary{intersections}
\usepackage[single=true]{acro}
\usepackage{glossaries}
\usepackage{upgreek}
\usepackage{grffile}
\usepackage{amssymb}
\usepackage{hyperref}
\newcommand{\sdown}{\ensuremath{_\textsc{sd}}}
\newcommand{\al}{\ensuremath{_\textsc{a}}}
\newcommand{\bo}{\ensuremath{_\textsc{b}}}

\newcommand{\Bf}{\ensuremath{_\textsc{bf}}}

\newcommand{\ls}{LS~5039\xspace}
\newcommand{\lsI}{LS~I+61$^\circ$303\xspace}

\let\glsa\glsuseri 
\newglossaryentry{grbs}{name={gamma-ray binary system},user1={Gamma-Ray Binary System},description={binary system with SED peaing in the gamma-ray energy band}}
\newglossaryentry{x}{name={X-ray},description={emission in the X-ray energy band}}
\newglossaryentry{g}{name={gamma ray},user1={gamma-ray},description={emission in the gamma-ray energy band}}
\newglossaryentry{lc}{name={light curve},description={general physical term}}
\newglossaryentry{lf}{name={Lorentz factor},description={general physical term}}
\newglossaryentry{suz}{name={Suzaku},description={satellite}}
\newglossaryentry{alf}{name={Alfv\'en},description={Hannes Alfv\'en}}

\DeclareAcronym{mev}{
  short = MeV ,
  long  =  megaelectronvolt; $10^6$~eV, 
  class = units ,
  first-style =reversed
}
\DeclareAcronym{gev}{
  short = GeV ,
  long  =  gigaelectronvolt; $10^9$~eV, 
  class = units ,
  first-style =reversed
}
\DeclareAcronym{tev}{
  short = TeV ,
  long  =  teraelectronvolt; $10^{12}$~eV, 
  class = units ,
  first-style =reversed
}
\DeclareAcronym{agn}{
  short = AGN ,
  long  =  active galaxy nucleus, 
  long-plural-form  =  active galaxy nuclei,
  class = units ,
}
\DeclareAcronym{hxd}{
  short = HXD ,
  long  =  hard \gls{x} detector ,
  class = instument ,
}

\DeclareAcronym{nus}{
  short = {\it NuSTAR} ,
  long = Nuclear Spectroscopic Telescope Array ,
  class = instument ,
  first-style =reversed ,
}
\newacronym{mev}{MeV}{megaelectronvolt, $10^6$~eV}
\newacronym{gev}{GeV}{gigaelectronvolt, $10^9$~eV}
\newacronym{tev}{TeV}{teraelectronvolt, $10^{12}$~eV}
\newacronym{nus}{{\it NuSTAR}}{Nuclear Spectroscopic Telescope Array}

\begin{document}

\title{
Sign of hard X-ray pulsation from the gamma-ray binary system LS 5039
}

\author{H. Yoneda}
\affiliation{Department of Physics, The University of Tokyo, 7-3-1 Hongo, Bunkyo, Tokyo 113-0033, Japan}
\affiliation{Kavli Institute for the Physics and Mathematics of the Universe (WPI), University of Tokyo, Kashiwa, Chiba 277-8583, Japan}
\affiliation{RIKEN Nishina Center, 2-1 Hirosawa, Wako, Saitama 351-0198, Japan}
\author{K. Makishima}
\affiliation{Kavli Institute for the Physics and Mathematics of the Universe (WPI), University of Tokyo, Kashiwa, Chiba 277-8583, Japan}
\affiliation{Department of Physics, The University of Tokyo, 7-3-1 Hongo, Bunkyo, Tokyo 113-0033, Japan}
\author{T. Enoto}
\affiliation{Extreme natural phenomena RIKEN Hakubi Research Team, Cluster for Pioneering Research, RIKEN, Hirosawa 2-1, Wako, Saitama, 351-0198, Japan}
\author{D. Khangulyan}
\affiliation{Department of Physics, Rikkyo University, 3-34-1 Nishi Ikebukuro, Toshima, Tokyo 171-8501, Japan}
\author{T. Matsumoto}
\affiliation{Department of Physics, The University of Tokyo, 7-3-1 Hongo, Bunkyo, Tokyo 113-0033, Japan}
\author{T. Takahashi}
\affiliation{Kavli Institute for the Physics and Mathematics of the Universe (WPI), University of Tokyo, Kashiwa, Chiba 277-8583, Japan}
\affiliation{Department of Physics, The University of Tokyo, 7-3-1 Hongo, Bunkyo, Tokyo 113-0033, Japan}

\begin{large}

\date{\today}
\begin{abstract}
To understand the nature of the brightest gamma-ray 
binary system LS 5039, hard X-ray data of the object,
taken with the {\it Suzaku} and {\it NuSTAR} observatories
in 2007 and 2016, respectively, were analyzed.
The two data sets jointly gave tentative evidence 
for a hard X-ray periodicity, with a period of $\sim 9$~s
and a period increase rate by $\sim 3 \times 10^{-10}$ s s$^{-1}$.
Therefore, the compact object in \ls is inferred to be
a rotating neutron star, rather than a black hole.
Furthermore, several lines of arguments suggest 
that this object has a magnetic field of 
several times \(\sim10^{10}~\rm T\),
two orders of magnitude higher than those of typical neutron stars.
The object is hence
suggested to be a magnetar, 
which would be the first to be found in a binary.
The results also suggest
that the highly efficient particle acceleration process,
known to be operating in \ls,
emerges through interactions between 
dense stellar winds from the massive primary star,
and ultra-strong magnetic fields of the magnetar.
\end{abstract}

\maketitle

{\em Introduction.}---
\Glspl{grbs} are a recently established, yet rare, 
class of astronomical objects \cite{Dubus2013}.
Their spectral energy distributions, peaked at above MeV energies, 
indicate extremely effective particle acceleration.
Although most of the proposed scenarios claim 
that they contain a non-accreting neutron star (NS) 
or an accreting black hole \cite{Mirabel2012},
the answer has been unknown in most cases,
let alone the detailed mechanism of their high-energy activity.

\ls is the brightest \gls{grbs} in the Galaxy,
consisting of an O-type primary star with a mass of $23 M_\odot$
and a compact secondary 
of unknown nature \cite{casarespossible2005}.
Its emission extends up to TeV energies, 
with a high bolometric luminosity 
of $\sim 1 \times $$10^{29}$ W \cite{Dubus2013, collmarls2014}.
It is suggested \cite{takahashistudy2009, khangulyan2008}
that particles are accelerated in this source
with exceptionally high efficiency,
e.g., up to tens of TeV in a few seconds.

Past observations of \ls revealed remarkable reproductivity 
of the soft X-ray orbital light curve \cite{kishishitalongterm2009}
and strong orbital-phase dependence in the \glsa{g} spectrum \cite{hess2006,fermi2009}.
These results disfavor the accretion scenario,
and suggest that LS 5039 harbors an NS. 
A plausible view \cite{dubus2015, takata2014} was 
that it contains a rotation-powered pulsar,
and its relativistic winds collide with the primary's stellar winds and then shock acceleration takes place.
This picture can explain the spectrum in the X-rays and GeV/TeV bands,
but cannot reproduce
the dominant component in MeV gamma-rays \cite{collmarls2014}.
Therefore, the compact star in \ls might not be an ordinary pulsar.
It could alternatively be a magnetar,
i.e., an NS with two orders of magnitude 
higher magnetic fields than typical pulsars,
as suggested for another gamma-ray binary system \lsI
based on its magnetar-like X-ray flare \cite{torresmagnetarlike2012}.

To confirm the presence of an NS (including a magnetar) in \ls,
detection of periodical pulses would be crucial.
Although such attempts using radio \cite{virginiamcswainradio2011} 
and soft X-rays \cite{readeep2011} have failed so far,
hard X-rays will be a better probe since they are less affected by the primary's stellar winds.
We hence performed a pulsation search from \ls,
for the first time in hard X-rays.

{\em Observation.}---
\ls was observed with the \ac{hxd} \cite{HXD2007}
onboard {\it Suzaku} \cite{suzaku2007}
from 2007 September 9 to 15
for a gross exposure of 497 ks (net 181 ks), 
covering about 1.5  orbital cycles \cite{kishishitalongterm2009}.
The \ac{hxd} was operated in  the normal mode 
with a time resolution of 61 $\upmu$s.
We utilize the events in 10--30 keV,
with a total number of $8.2\times10^4$,
of which about 90\% are background.

Nine years later, {\it NuSTAR} \cite{nustar2013} observed \ls from 2016 September 1 to 5
for a gross exposure of 346 ks (net 191 ks), 
or about one orbital cycle.
We extracted 10--30 keV events
from a circular region centered at the source 
with a radius of 30 arcsec.
The total number of events is  $1.2\times10^4$ 
of which 4\% is background.
The time resolution is 10 $\upmu$s 
after correcting for the clock drifts.

{\em Method.}---
If the putative NS in \ls has
a mass of $\sim 1.4 M_\odot$,
its projected orbital radius should be
$\sim 50$ light sec.
Thus, individual pulses, suffering periodic changes 
by $\sim 50$ s in their arrival times, would be smeared out
unless the pulse period is $P_{\rm NS} \gg 50$ s.
Although this problem could be avoided
by correcting the photon arrival times
for the NS's binary motion,
the orbital solutions currently available from optical observations \cite{casarespossible2005, aragonaorbits2009, Sarty2011}
are not accurate enough for this purpose.

We thus search for pulsations, 
first without corrections for the NS's orbital motion.
Its line-of-sight velocity would cause the Doppler modulation up to
\begin{equation}
\Delta v /c \sim 50 ~ \mathrm{(s)}\times 2\pi/P_{\rm orb} = 1\times 10^{-3} ~.
\label{eq:0.001}
\end{equation}
A simple way to mitigate this $\pm 0.1\%$ period change 
is to divide the whole data into many subsets,
each with a duration of $\Delta T$
which is short enough to approximately satisfy 
\begin{equation}
P_{\rm NS} / \Delta T \gtrsim 1\times 10^{-3} ~.
\end{equation}
Then, the Fourier frequency resolution 
becomes no higher than Eq.(1). 
The power spectra thus calculated from individual subsets
are merged incoherently into an averaged spectrum with improved statistics.
Requiring each subset to include $\gtrsim 10$ source photons,
we limit our search to $P_{\rm NS}>1$ s.
Thus, we focus on slowly-rotating neutron stars
(Supplemental Material (SM) {\S}A \cite{suppl}).

{\em Results.}---
First, we Fourier-analysed the 10--30 keV HXD data 
over a frequency range of $10^{-2}$--1.0 Hz,
employing $\Delta T = 4096, 8192$, and 16384 s,
and obtained Figure~\ref{fig_stackedana_fft} (a).
The result for $\Delta T = 8192$ s reveals a clear peak 
at $P_{\rm NS}= 8.96$ s, where the power reaches 3.79 above the average of 2.0.
The post-trial probability of this peak was estimated as follows. 
As we used 55 data subsets when $\Delta T = 8192$ s,
each Fourier component in Figure~\ref{fig_stackedana_fft} (a)
obeys a $\chi^2$ distribution with 110 d.o.f. 
(SM {\S}B \cite{suppl})
and the local chance probability of the $8.96$ s peak 
becomes $4.5\times10^{-8}$.
Because 8192 independent frequencies were tested,
the chance probability considering look-elsewhere effects,
becomes ${\cal P}_{\rm ch}= 4.5\times10^{-8} \times 8192
= 3.7\times10^{-4}$.
Finally considering
a factor 3 
which is the numbers of $\Delta T$ tested,
we obtain ${\cal P}_{\rm ch}= 1.1\times10^{-3}$.
This estimation was also confirmed by a Monte-Carlo simulation.
Thus, the 8.96 s periodicity has
a confidence level close to 99.9\%.
The ratio $P_{\rm NS}/\Delta T$ = 0.11\% is fully self-consistent
within our framework; see Eq.(2).
Furthermore, this periodicity is unlikely to be due to
contamination from other sources (SM {\S}C \cite{suppl}).

The above result was further studied
using $\mathrm{Z^2}$ statistics \cite{deJager1989}, 
a test for weak periodic signals with unbinned likelihood evaluation. 
Using only the fundamental harmonic,
we calculated $\mathrm{Z^2}$ over $P_\mathrm{NS} = $ 1--100 s,
also from individual subsets of length $\Delta T$,
and incoherently stacked the results into a single $\mathrm{Z^2}$ periodogram.
We changed $\Delta T$ 
from 3000 s to 12000 s, with a step of 1000 s.
Figure~\ref{fig_stackedana_fft} (b-1) shows the case with $\Delta T = 5000$ s,
where the periodicity appears clearly at $P_{\rm NS} = 8.960 \pm 0.009$ s;
the quoted error is dominated by the orbital Doppler shifts.

The significance of this $\mathrm{Z}^2$ peak was evaluated with a Monte-Carlo method.
In a single trial, we generated the entire subsets,
each with the same photon counts and same observing windows
as the actual data, but with no intrinsic periodicity.
Each subset was $\mathrm{Z}^2$ analyzed.
The results were again stacked into a single periodogram 
covering $P_\mathrm{NS} =$ 1--100 s,
and the maximum $\mathrm{Z}^2$ was registered.
After $2\times10^4$ trials, we found 3 cases with the maximum $\mathrm{Z}^2$
higher than in Figure~\ref{fig_stackedana_fft} (b-1).
This yields ${\cal P}_{\rm ch}=1.5 \times 10^{-4}$ for the observed peak,
considering look-elsewhere effects
over the 1--100 s range.
Counting the 10 steps in $\Delta T$,
we finally obtain ${\cal P}_{\rm ch} = 1.5 \times 10^{-3}$
in agreement with the Fourier analysis.

Similarly, we analyzed the 10--30 keV {\it NuSTAR} data of \ls. 
First, we performed the same Fourier analysis,
using $\Delta T =$ 4096, 8192, and 16384 s,
but found no significant peaks in the averaged power spectra.
To search for weaker pulsations, 
the data were then analyzed with the $\mathrm{Z^2}$ statistics.
We again used only the fundamental harmonic,
and limited the search range to $P_{\rm NS}$ = 7--11 s,
which accommodates period changes at a rate of 
$|\dot{P}_{\rm NS}| < 7.0\times10^{-9}~\mathrm{s~s^{-1}}$,
from the value indicated by {\it Suzaku}.
We again changed $\Delta T$ from 3000 s to 12000 s, with a step of 1000 s.
When $\Delta T = 10000$ s,
the {\it NuSTAR} data yielded an obvious peak at 
$P_{\rm NS}=9.046\pm 0.009\mathrm{~s}$,
as in Figure~\ref{fig_stackedana_fft} (b-2).
Additionally, the $Z^2$ value took the maximum at 9.046 s regardless of $\Delta T$.

Again, the significance of this $\mathrm{Z}^2$ peak was Monte-Carlo evaluated.
After $2\times10^4$ trials, we found 62 cases with $\mathrm{Z}^2$
higher than in Figure~\ref{fig_stackedana_fft} (b-2).
This yields ${\cal P}_{\rm ch}=3.1 \times 10^{-3}$ for this peak, considering look-elsewhere effects
for the limited period search range of 7--11 s.
Including the trials of $\Delta T$, 
we obtain ${\cal P}_{\rm ch} \lesssim 3.1 \times 10^{-2}$.

The two observations 9 years apart by the two satellites thus gave evidence for the $\sim 9$ s hard X-ray pulsation.
Therefore, the compact object in \ls is inferred to be an NS with a spin period of \(P_{\rm NS}\sim 9\)~s,
and its derivative of \(\dot{P}_{\rm NS}\sim 3\times 10^{-10}\)~s~s\({}^{-1}\).

Finally, to confirm that the pulsed emission 
really comes from the compact object in \ls,
we repeated the $\mathrm{Z^2}$ analysis
separately using the entire {\it Suzaku} and {\it NuSTAR} data sets,
subdividing neither of them, but correcting the photon arrival times for the orbital motion.
The orbital period was fixed at the 
optical value of $P_{\rm orb}=3.90608$ days \cite{aragonaorbits2009}.
We first scanned the orbital parameters
over the ranges given in Table~\ref{tab_binaryDemoSuzakuparameter1}.
After finding a $\mathrm{Z^2}$ maximum 
(separately for the two data sets),
finer search steps were employed.
We used the harmonics up to $m=4$ 
to describe the pulse-profile details.

The {\it Suzaku} data yielded 
the orbital parameters as shown in Table~\ref{tab_binaryDemoSuzakuparameter1}.
They are consistent with the optical information.
As the {\it NuSTAR} data gave several orbital solutions
with comparable significance,
Table~\ref{tab_binaryDemoSuzakuparameter1} lists
the one that is closest to the {\it Suzaku} solution (SM {\S}E \cite{suppl}).
The {\it Suzaku} and {\it NuSTAR} pulse profiles,
derived with the respective orbital solutions,
are shown in Figure~\ref{fig_profile}; 
they exhibit similar three peaks with separations of $\sim 0.25$ pulse phases.

Figure~\ref{fig_periodgram} shows the $\mathrm{Z^2}$ periodograms
after applying the orbital solutions in Table~\ref{tab_binaryDemoSuzakuparameter1}.
Both the {\it Suzaku} and {\it NuSTAR} data show clear peaks, 
with pre-trial probabilities of $1.7\times10^{-11}$ and $7.3\times10^{-11}$, 
respectively. Considering look-elsewhere effects,
these yield ${\cal P}_{\rm ch}= 1.2\times10^{-2}$ ({\it Suzaku})
and ${\cal P}_{\rm ch}= 7.0\times10^{-2}$ ({\it NuSTAR}),
since the number of independent trials in the parameter search 
was Monte-Carlo estimated as $7.2\times 10^{8}$ and $9.6\times 10^{8}$, respectively.
For further examination,
we scanned $P_{\rm orb}$,
to find that the pulse significance becomes maximum
at $P_{\rm orb} =3.90 \pm 0.05$ ({\it Suzaku})
and $3.94 \pm 0.08$ days ({\it NuSTAR}) in a good agreement with the reported value.

Although we have thus obtained signs of pulsation from \ls,
we are still left with several problems.
(1) The orbital solutions from the two data 
disagree on some parameters beyond their statistical errors.
(2) The orbital corrections did not increase 
the pulse significance as large as expected.
(3) After background subtraction,
the two data give 
rather different pulse fractions (Figure~\ref{fig_profile}),
$0.68\pm0.14$ ({\it Suzaku}) and $0.135\pm0.043$ ({\it NuSTAR}).
The 10--30 keV fluxes were also different from each other,
$(10.7\pm1.2) \times 10^{-12}~\mathrm{erg~cm^{-2}~s^{-1}}$ ({\it Suzaku})
and
$(8.26\pm0.10) \times 10^{-12}~\mathrm{erg~cm^{-2}~s^{-1}}$ ({\it NuSTAR}).

To find clues to these issues,
we subdivided the {\it Suzaku} and {\it NuSTAR} data into several segments of similar lengths,
and folded each with the orbital solution.
Although these segments reproduced the pulse profile globally,
its fine structures varied (SM {\S}F \cite{suppl}).
Thus, the pulses could suffer from some modulation besides the orbital motion, which may have biased the orbital parameters beyond statistical errors.  

We also analyzed the 3--10 keV {\it NuSTAR} data
using the parameters in Table~\ref{tab_binaryDemoSuzakuparameter1}, 
but found no periodic signals at the period.
The upper-limit pulse fraction is 2.8\%  (99\% confidence level, SM {\S}G \cite{suppl}).

{\em Discussion.}---
We have obtained evidence for a $\sim 9$ s pulsation from \ls.
Moreover, the mass of this pulse emitter 
is constrained as $1.23$--$2.35~M_\odot$ (SM {\S}A \cite{suppl}).
We therefore infer that \ls harbors an NS; 
this would settle the controversy as to the compact object 
in this system.

Our results also provide clues to the energy source
that powers the remarkable non-thermal processes of \ls.
Generally, we can think of four candidates;
rotational energy of the NS, 
stellar winds from the massive star, 
gravitational energy due to mass accretion, 
and magnetic energy of the NS.
From the detected $P_{\rm NS}$ and its derivative $\dot{P}_{\rm NS}$,
the first one is excluded immediately,
because the spindown energy loss of the NS, 
$
  L\sdown = (2\uppi)^2 I\dot{P}_{\rm NS}/P_{\rm NS}^3 \sim10^{27}\rm\,W\ 
$
with $I$ the NS moment of inertia, is much below
the observed luminosity of $\sim 10^{29}$ W.
The same applies to the second candidate, 
because the  power available in this case
is $< 10^{25}$ W (SM {\S}H \cite{suppl}). 
The third candidate is also excluded, 
because neither the MeV-peaked spectrum 
nor the largely positive $\dot{P}_{\rm NS}$ of \ls 
is observed from accreting pulsars.
The good reproducibility of the soft X-ray orbital light curve
also disfavors the accretion scenario \cite{kishishitalongterm2009}.

Hence the sole option is left:
dissipation of the NS's magnetic energy,
of which the available output is given as
\begin{equation}
\begin{split}
 L\Bf &= \frac{B_\mathrm{NS}^2 R_\mathrm{NS}^3}{6\tau}\\
 &\sim 10^{30} \left(\frac{B_\mathrm{NS}}{10^{11}\mathrm{T}}\right)^2 
   \left(\frac{\tau}{500\mathrm{yr}}\right)^{-1}\rm~W ~,
   \end{split}
\end{equation}
where $R_\mathrm{NS}\approx 10$ km, $B_\mathrm{NS}$,
and $\tau=\frac{1}{2}P_{\rm NS}/\dot{P}_{\rm NS} \sim 500$ yr 
denote the NS's radius,
surface magnetic field,
and the characteristic age, respectively.
Then, the need for $L\Bf>10^{29}$ W is satisfied if 
 \(B_\mathrm{NS} \gtrsim 3\times10^{10}\rm\,T\).
This $B_\mathrm{NS}$ corresponds to those of  magnetars, 
whose radiation is powered by their magnetic energies.
This scenario is reinforced by the detected  $P_{\rm NS} \sim 9$ s,
which is typical of magnetars \cite{enoto2010}.
Additionally, the sporadic variations in the pulse properties, 
suggested by the present data,
are also observed from magnetars \cite{Wood2004, makishimapossible2014}.

The magnetar scenario can also explain the absence of accretion in \ls.
In these binaries, the ram pressure of the 
primary's stellar winds is balanced 
by the NS's magnetic pressure, at a radius 
\begin{equation}
\begin{split}
R\al&=
\frac{R_{\rm NS} \left( B_{\mathrm{NS}} a_{\mathrm{x}} \right)^{1/3}}
            {( 2 \dot{M}_\mathrm{w} v_\mathrm{w} )^{1/6}}\\
 &\sim 2\times 10^{8}  \left(\frac{B_{\mathrm{NS}}}{10^{11}~\mathrm{T}}\right)^{1/3} 
~\mathrm{m}\,,
\end{split}
\end{equation}
where $a_{\mathrm{x}}\sim 50$ light-sec is the binary separation,
$\dot{M}_\mathrm{w} = 10^{-6}M_{\odot}~\mathrm{yr}^{-1}$
is a typical wind mass-loss rate of O stars,
and $v_\mathrm{w} = 2000~\mathrm{km~s^{-1}}$ is the typical wind velocity.
For $B_\mathrm{NS} \gtrsim 2 \times 10^{10}$ T,
this ${R_{\rm A}}$ exceeds the Bondi-Hoyle capture radius 
$R\bo \sim 1\times10^{8} ~\mathrm{m}$:
the strong magnetic pressure suppresses
the gravitational wind capture.

As already suggested for another system \cite{torresmagnetarlike2012},
we thus propose that \ls harbors a magnetar,
with $B_{\rm NS}\sim$ several times $10^{10}$ T.
For reference, 
$B_{\rm NS} \sim 10^{11}~\mathrm{T}$ is also derived
assuming the object to spin down
via magnetic dipole radiation,
but this should be taken as an upper limit 
since the spindown would be accelerated 
by interactions with the stellar winds.

Compared to \ls,
isolated magnetars are less luminous ($\sim 10^{28}$ W),
and lack the strong MeV component \cite{enoto2011}.
These differences may be attributed to 
the uniqueness of \ls  
that it is immersed in the dense stellar winds.
We hence suggest that
interactions of the strong magnetic field with the stellar winds 
are essential, both in generating 
the distinctive non-thermal emission
and enhancing its magnetic energy dissipation
to afford a higher luminosity.
A plausible scenario is that the stellar wind pressure 
rearranges the magnetar's magnetosphere 
and its magnetic energy is converted efficiently 
into radiation via,  e.g., magnetic reconnection.
TeV electrons would be produced efficiently in the reconnection region.
Possibly, the MeV emission is also produced
therein, via synchrotron radiation in the strong magnetic fields.

In closing, 
we found a periodicity around 9 s from the {\it Suzaku} hard X-ray data with a post-trial probability of $1.1\times10^{-3}$. 
Moreover, the {\it NuSTAR} data gave tentative evidence of the 9 s pulsation. 
These data jointly suggest $\dot{P}_{NS} 
\sim 3\times10^{-10}~\mathrm{s~s^{-1}}$,
although further confirmation of these values is needed
since the {\it NuSTAR} detection is not significant by itself.
With this reservation,
our result is thought 
to have the three-fold significance.
(1) It casts light on a novel scenario for 
the highly efficient acceleration mechanism in astrophysics.
(2) The NS in \ls, the first magnetar candidate found in a binary,
will provide clues to the birth and evolution of magnetars.
(3) Refining the mass constraint through future observations,
we may possibly tell whether the mass range of magnetars
is different from that of ordinary NSs.

\begin{acknowledgments}
H.Y. acknowledges the support of the Advanced Leading Graduate Course for Photon Science (ALPS).
We acknowledge support from JSPS KAKENHI grant numbers 16H02170, 16H02198, 17J04145,  18H01246, 18H05463, 18H03722, 18K03694, and 20H00153.
\end{acknowledgments}

\begin{figure*}[b]

    \begin{center}
    \includegraphics[width = 16 cm]{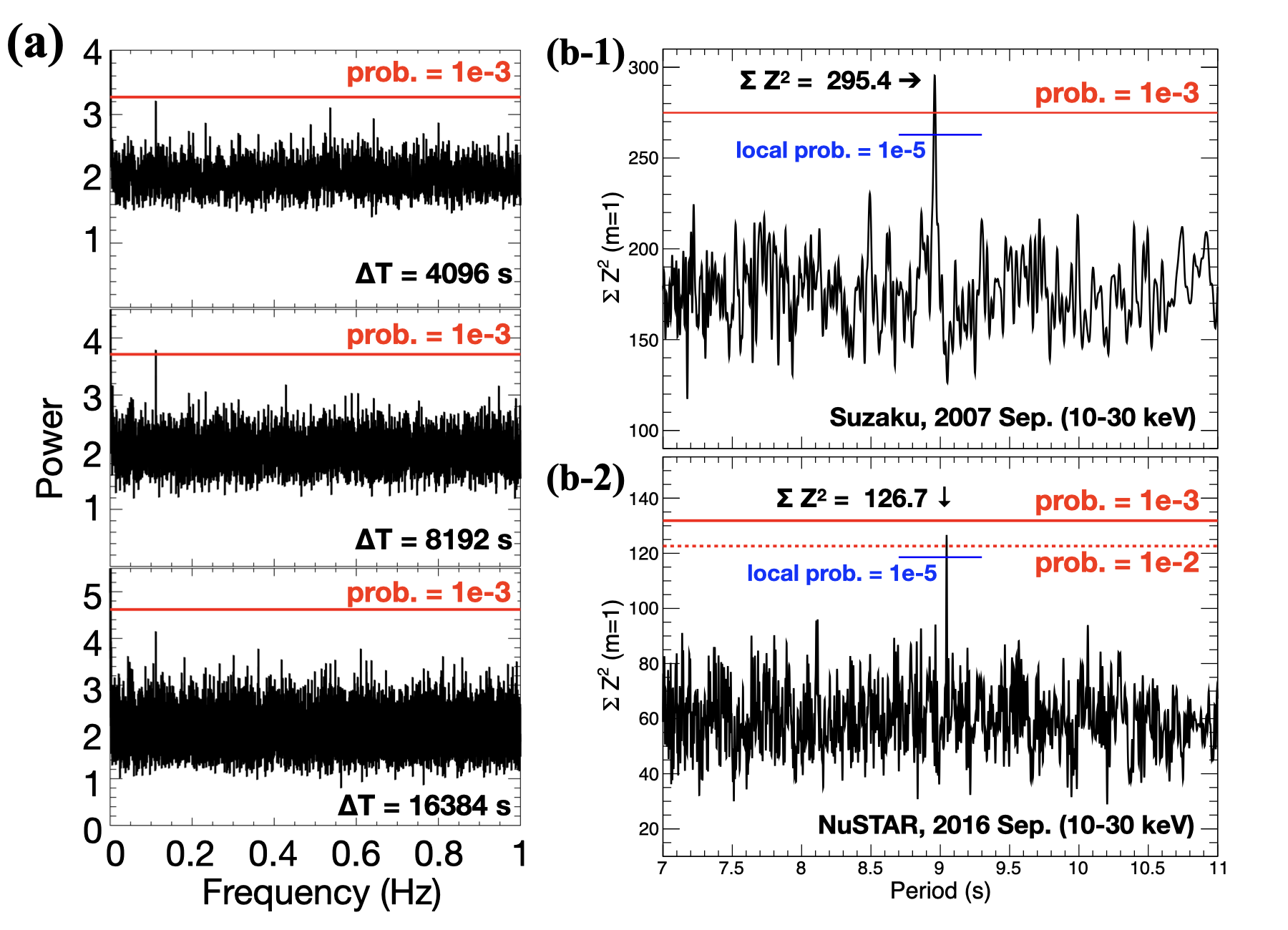}
    \end{center}

\caption{\large{Evidence of the pulsation from \ls.
(a) 
Fourier analysis of the 10--30 keV  {\it Suzaku} data, 
where red lines indicate signal strengths 
that arise with a post-trial probability of $1.0\times10^{-3}$
when considering the look-elsewhere effect.
The data are divided into subsets with $\Delta T=$ 4,096, 8,192, and 16,384 s,
and the power spectra from individual subsets are averaged. 
(b)
$\mathrm{Z^2}$ periodograms in 10--30 keV, 
shown for $P_{\rm NS}=7$--11 s.
Panel (b-1) shows the {\it Suzaku} result
(though the search was conducted over 1--100 s), 
averaged over 86 subsets with $\Delta T=5,000$ s,
and panel (b-2) that of {\it NuSTAR}
from 30 subsets with $\Delta T = 10,000$ s.}
}
\label{fig_stackedana_fft}

\end{figure*}

\begin{figure*}[b]

    \begin{center}
    \includegraphics[width = 8 cm]{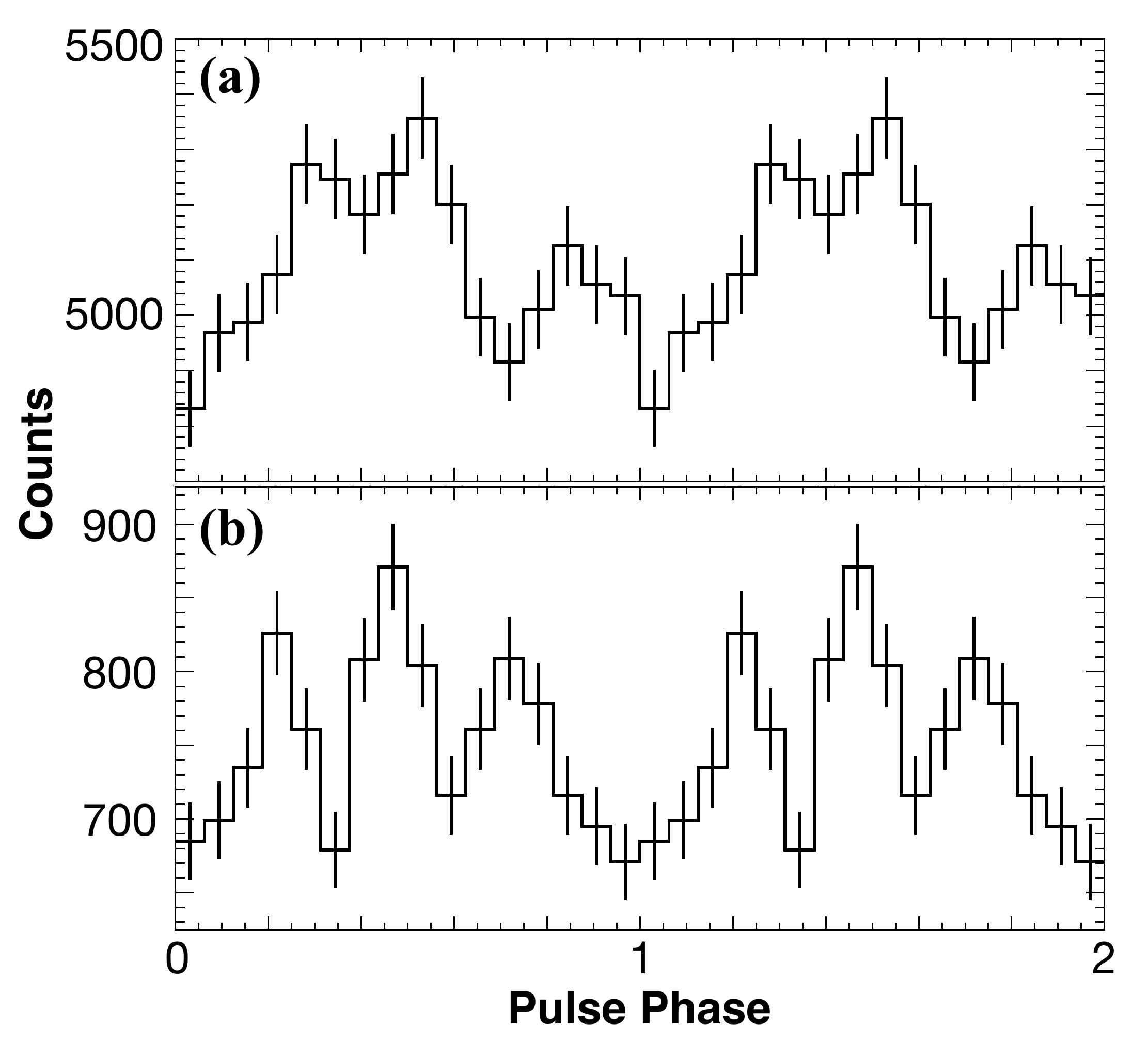}
    \end{center}
    
\caption{\large{
The 10--30 keV folded pulse profiles 
from {\it Suzaku} (panel a) and {\it NuSTAR} (panel b),
obtained using the best-estimated orbital parameters 
and optimum $P_{\rm NS}$ of the respective observations
(Table~\ref{tab_binaryDemoSuzakuparameter1}).
}}
\label{fig_profile}
\end{figure*}

\begin{figure*}[b]

    \begin{center}
    \includegraphics[width = 15 cm]{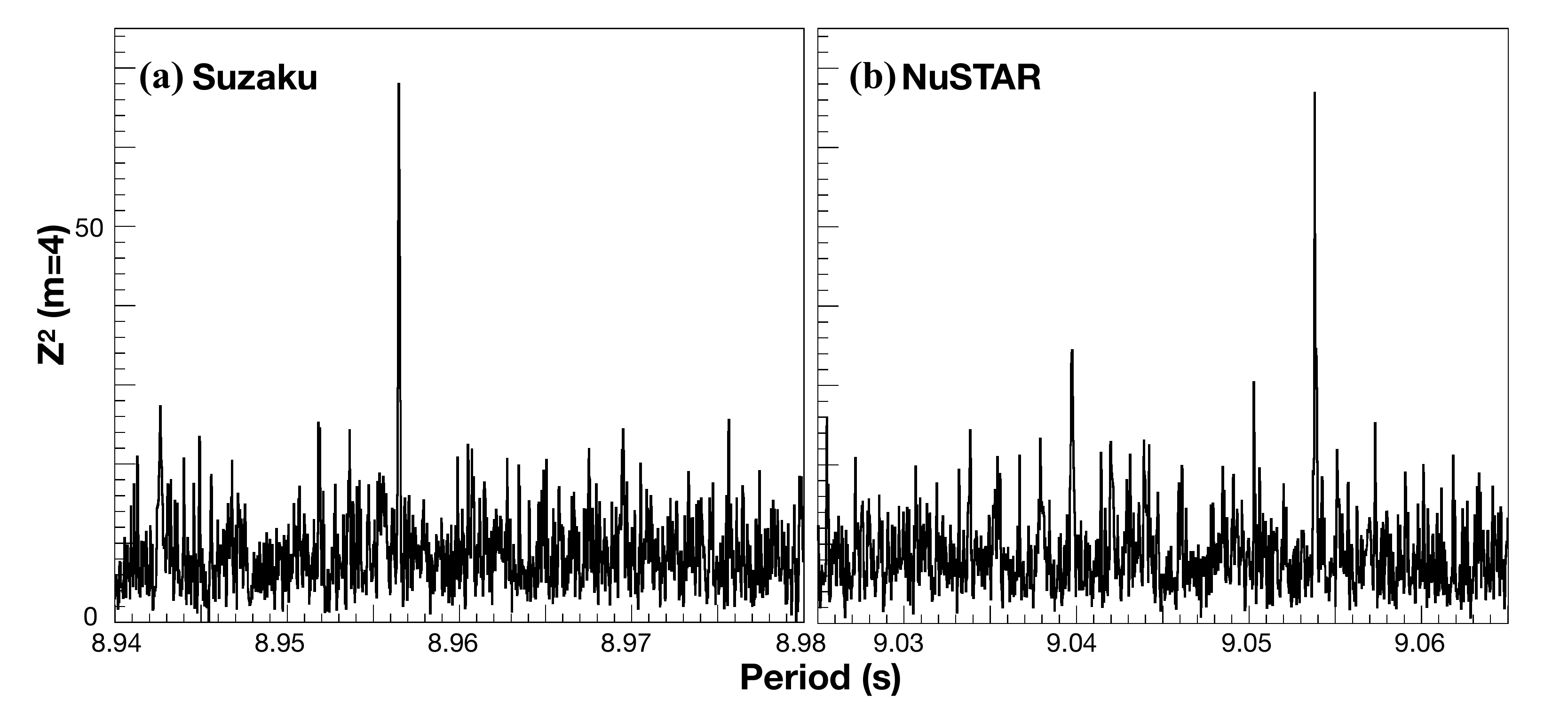}
    \end{center}
    
\caption{\large{$\mathrm{Z^2}$ periodograms calculated coherently
using the entire data of {\it Suzaku} (panel a) and {\it NuSTAR} (panel b),
after the orbital-motion corrections 
using the  best-fit  parameters in Table~\ref{tab_binaryDemoSuzakuparameter1}.
Here up to the 4th harmonics are incorporated.
}}
\label{fig_periodgram}
\end{figure*}

\begin{table*}[b]
\caption{\large{The search ranges of the orbital parameters and the best-fit results,
obtained from {\it Suzaku}/{\it NuSTAR} observations. 
Errors of the best-fit values refer to 90\% confidence limits}.}
\label{tab_binaryDemoSuzakuparameter1}
\centering
\medskip
\large
\begin{tabular}{ccccccc} \hline
parameter & \multicolumn{2}{c}{search range} & \multicolumn{2}{c}{best-fitting value} & \multicolumn{2}{c}{optical results} \\
& min & max & {\it Suzaku} & {\it NuSTAR} & \cite{aragonaorbits2009} & \cite{Sarty2011}
\\ \hline
$a_\mathrm{x} \sin \theta$ [light sec.] & $30.0$ & $67.375$ & $53.05^{+0.70}_{-0.55}$ & $48.1^{+0.4}_{-0.4}$ & - & -\\
$e$ & $0.16$ & $0.39$ & $0.278^{+0.014}_{-0.023}$ & $0.306^{+0.015}_{-0.013}$ & $0.337 \pm 0.036$ & $0.24 \pm 0.08$\\
$\omega$ [deg.] & $45$ & $65$ & $54.6^{+5.1}_{-3.3}$ & $56.8^{+2.3}_{-3.1}$ & $56.0 \pm 5.8$ &  $57.3 \pm 21.8$\\
$\tau_0$ ({\it Suzaku})$^{*}$ & $-0.05$ & $0.06875$  & $0.067^{+0.009}_{-0.012}$ & - & $-0.022 \pm 0.017$ &  $0.030 \pm 0.07$\\
$\tau_0$ ({\it NuSTAR})$^{*}$ & $0.5$ & $0.74875$ & - & $0.7285^{+0.0078}_{-0.0058}$ & $0.546 \pm 0.034$ &  $0.615 \pm 0.33$\\
$P_{\rm NS}$ [s] ({\it Suzaku})& $8.94$ & $8.98$ & $8.95648(4)$ & - & - & -\\ 
$P_{\rm NS}$ [s] ({\it NuSTAR})& $9.025$ & $9.065$ & - & $9.05381(3)$ & - & -\\ \hline
$\mathrm{Z^2} (m=4)$ & - & - & $67.97$ & $66.87$ & - & -\\ \hline
\end{tabular}
\begin{itemize}
\item[] $^{*}$: The orbital phase at the time of the initial event in the data. The time of the initial event of {\it Suzaku} and {\it NuSTAR} is 54352.7163 
and 57632.0952 (MJD TT) respectively.
\end{itemize}
\end{table*}

\end{large}

\end{document}